\documentclass[%
  aip,
  jcp,
  citeautoscript,
  superscriptaddress,
  amsmath,
  amssymb,
  reprint
]{revtex4-2}

\usepackage[version=4]{mhchem}
\usepackage{graphicx}
\usepackage[dvipsnames]{xcolor}
\usepackage[colorlinks=true,
            linkcolor=Blue,
            citecolor=Blue,
            filecolor=Blue,
            urlcolor=Blue
            ]{hyperref}
\usepackage{layouts}
\usepackage{multirow}

\graphicspath{{./figs/}}

\begin{document}

\title{Single-atom dopants in plasmonic nanocatalysts}

\author{Daniel Sorvisto}
\email{daniel.sorvisto@aalto.fi}

\author{Patrick Rinke}
\email{patrick.rinke@aalto.fi}

\author{Tuomas P. Rossi}
\email{tuomas.rossi@aalto.fi}
\affiliation{%
    Department of Applied Physics,
    Aalto University,
    Espoo, Finland
}

\date{\today}

\begin{abstract}
Bimetallic nanostructures combining plasmonic and catalytic metals are promising for tailoring and enhancing plasmonic hot-carrier generation utilized in plasmonic catalysis.
In this work, we study the plasmonic hot-carrier generation in noble metal nanoparticles (Ag, Au, Cu) with single-atom dopants (Ag, Au, Cu, Pd, Pt) with first-principles time-dependent density-functional theory calculations.
Our results show that the local hot-carrier generation at the dopant atom is significantly altered by the dopant element while the plasmonic response of the nanoparticle as a whole is not significantly affected.
In particular, hot holes at the dopant atom originate from the discrete d-electron states of the dopant.
The energies of these d-electron states, and hence those of the hot holes, depend on the dopant element, which opens up the possibility to tune hot-carrier generation with suitable dopants.
\end{abstract}


\maketitle

\section{Introduction}

Chemical reactions are a fundamental transformation and creation process for substances and materials. The rates of many artificial and naturally occurring chemical reactions can be increased using catalysts. \cite{Rod14}
A promising driver of catalytic reactions is the plasmonic response of metal nanoparticles (NPs). \cite{AslRaoCha18}
Light can excite the free electrons of a NP to coherently oscillate in form of a collective localized surface plasmon resonance (LSPR). \cite{KreVol95}
When the excited LSPR subsequently decays through Landau damping, \cite{YanBro92} it results in a non-thermal distribution of electrons and holes, so-called hot carriers, \cite{BroHalNor15} which can trigger chemical reactions in surrounding molecules. \cite{LinAslBoe15}

The plasmonic-catalytic performance of metal NPs can be enhanced with antenna-reactor designs \cite{SweZhaZho16}, in which a plasmonic metal (such as Ag, Au, Cu) is mixed with a catalytic metal (such as Pd, Pt). \cite{SytVadDio19}
The components of such bimetallic structures could be either separated or alloyed, and at the extreme limit are single-atom alloys \cite{HanGiaFly20}, in which single dopant atoms are embedded into a host metal NP.
In the context of plasmonic catalysis, such single-atom reactor sites in a plasmonic NP have been experimentally demonstrated to provide high efficiencies and selectivities, \cite{ZhoMarFin20} while also keeping the needed amount of the potentially expensive catalytic metal low.
However, it can be experimentally challenging and time-consuming to reveal underlying mechanisms and explore the vast materials space of potential compounds.
Theoretical and computational methods could come to the rescue. They have a proven track record in modeling plasmonic hot-carrier generation \cite{SunNarJer14, BerMusNea15, ForRanLis18,  DouBerFra19, BerDouSan20, RosErhKui20, JinKahPap22} and transfer \cite{DouBerSan16, KumRosMar19, MaGao19, DouBoxMau21, MaWanGao22, FojRosKui22} and in designing bimetallic compounds. \cite{RanForLis18}

\begin{figure}[t]
    \centering
    \includegraphics[scale=1]{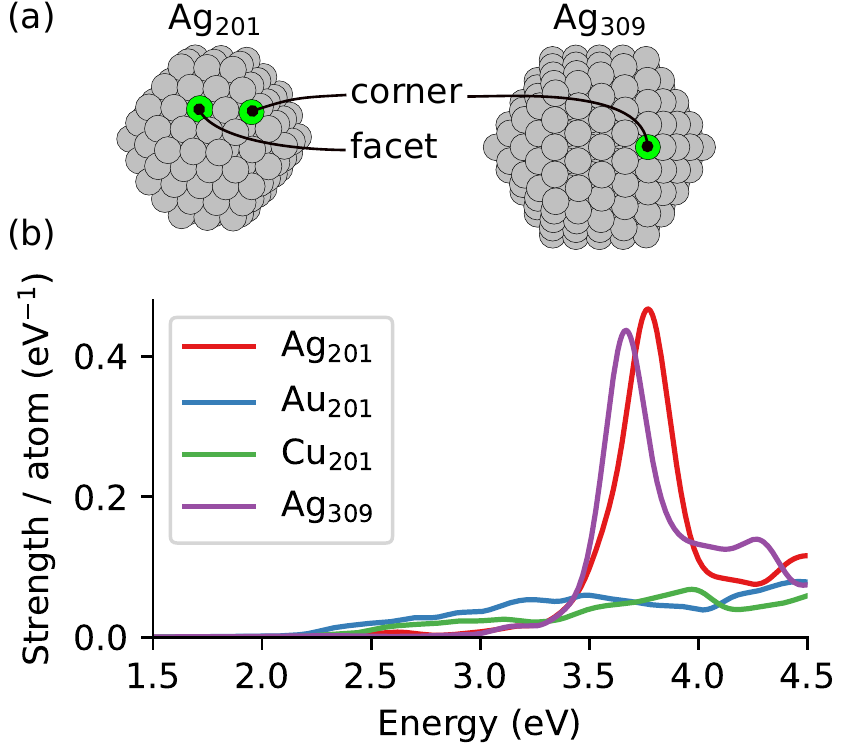}
    \caption{%
        (a)
        Atomic structure of regularly truncated octahedral \ce{Ag201} NP and icosahedral \ce{Ag309} NP with considered corner and \{111\} facet atom sites indicated.  \ce{Au201} and \ce{Cu201} are similar to \ce{Ag201}.
        (b)
        Photoabsorption spectra of the bare NPs presented as the dipole strength functions.
    }
    \label{fig:system}
\end{figure}

In this work, we model the plasmonic-catalytic properties of prototypical plasmonic NPs with single-atom dopants to elucidate how single dopants affect the plasmon-generated hot carriers and how the electronic structure of the dopant site is reflected in the carrier distributions.
We consider the octahedral plasmonic NPs \ce{Ag201}, \ce{Au201}, and \ce{Cu201}, in which a single atom on a corner or  a \{111\} facet site on the surface is replaced by Ag, Au, Cu, Pd, or Pt (\autoref{fig:system}a).
In addition, we consider a larger icosahedral \ce{Ag309} NP with a single corner atom dopant to estimate the effect of particle size and shape.

The paper is organized as follows: In \autoref{sec:methods}, we describe the calculation method based on density-functional theory (DFT) and time-dependent DFT (TDDFT) and the details of the simulations.
In \autoref{sec:results}, we present the results of plasmonic-catalytic response and analyse the contribution of the dopants.
We summarize the work with an outlook in \autoref{sec:conclusions}.

%%%% FIGURE %%%%%%%%%%%%%%%%%%%%%%%%%%%%%%%%%%%%%%%%%%%%%%%%%%%%%%%%%%%%%%%%%%
%%%% FIGURE %%%%%%%%%%%%%%%%%%%%%%%%%%%%%%%%%%%%%%%%%%%%%%%%%%%%%%%%%%%%%%%%%%

\section{Methods}
\label{sec:methods}

The ground-state electronic structure of the NPs is modeled with Kohn--Sham (KS) DFT \cite{HohKoh64, KohSha65} as implemented in the open-source GPAW code package \cite{EnkRosMor10} in localized basis sets mode \cite{LarVanMor09}.
The geometries are relaxed with the Perdew--Burke--Ernzerhof (PBE) \cite{PerBurErn96} exchange-correlation functional and the local Broyden--Fletcher--Goldfarb--Shanno (BFGS) optimizer in the open-source Atomic Simulation Environment (ASE) package \cite{LarMorBlo17} in a three-step procedure: first the plain NP (e.g. \ce{Ag201}) is relaxed, then a single atom is replaced, after which the structure is relaxed again to obtain the final structure.
Atomic forces are converged to be below 0.01 eV\AA$^{-1}$.

Before calculating the plasmonic response, the ground-state electron density is recalculated for fixed atomic positions with the solid-state modification of the Gritsenko--van Leeuwen--van Lenthe--Baerends exchange-correlation potential (GLLB-sc), \cite{GriLeeLen95, KuiOjaEnk10}
which improves the description of the d-electron states in noble metals in comparison to PBE. \cite{KuiSakRos15, YanJacThy11, YanJacThy12}
Then, linear impulse response and photoabsorption cross sections are calculated with TDDFT \cite{RunGro84} and the real-time propagation TDDFT implementation \cite{KuiSakRos15} in GPAW with the adiabatic GLLB-sc kernel \cite{KuiSakRos15} and the $\delta$-kick technique \cite{YabBer96}.

The hot-carrier distributions are calculated with the convolution method developed in Ref.~\onlinecite{RosErhKui20}: in short, we analyze the linear response to a Gaussian electric field pulse of the form $\mathcal{E}_{0} \cos\big(\omega_0 (t - t_0)\big) \exp\big(-(t - t_0)^2/\tau_0^2\big)$ with $\mathcal{E}_{0}=51\text{\,\textmu{}V/\AA}$, $t_0 = 10\text{\,fs}$, and $\tau_0 = 3\text{\,fs}$, and evaluate the hot-carrier distributions at $t=30\text{\,fs}$ from the pulse response.
The pulse frequency $\omega_0$ is set to 3.77\,eV for \ce{Ag201}, \ce{Au201}, and \ce{Cu201} and 3.67\,eV for \ce{Ag309}, corresponding to the LSPR frequency of the Ag particles (\autoref{fig:system}b).
All 201-atom particles use the same pulse frequency for consistency. 
The atom-projected quantities are obtained by Voronoi decomposition as described in Ref.~\onlinecite{RosErhKui20}.

Hot carrier occupation probabilities scale linearly with the pulse strength as long as the linear response regime holds. With the weak pulse strengths employed in this work, we achieve occupation probabilities on the order of $10^{-8}\,\text{eV}^{-1}$.

The perturbing electric field pulse is aligned parallel to the axis that connects the center of the particle with the dopant atom.
While the dopant atoms cause qualitatively similar effects with different pulse directions, the exact distributions depend on the pulse direction (shown for \ce{Ag201} with Pt dopant in Supplementary Fig.~S1).

We use projector augmented-wave (PAW) \cite{Blo94} sets with 11 valence electrons for Ag, Au, Cu, and 10 valence electrons for Pd and Pt, treating the remaining electrons as frozen core.
For the relaxation calculations, we use standard double-$\zeta$ polarized (dzp) basis sets, but the extended ``p-valence'' basis set is used for dynamic response calculations, similarly to Ref.~\onlinecite{RosErhKui20}.
In time propagation, we apply a time step of 10\,as and propagate for a total of 30\,fs. All spectra are broadened using Gaussian damping with $\sigma = 0.07\text{\,eV}$.
A real-space grid spacing parameter $h$ of 0.2\,\AA\ is used and the system is surrounded by a vacuum region of at least 6\,\AA.
The Hartree potential is evaluated with a Poisson solver using the monopole and dipole corrections for the potential.
Fermi-Dirac smearing of 0.05\,eV is applied to the occupation numbers to facilitate convergence.

\section{Results and Discussion}
\label{sec:results}

%%%% FIGURE %%%%%%%%%%%%%%%%%%%%%%%%%%%%%%%%%%%%%%%%%%%%%%%%%%%%%%%%%%%%%%%%%%
\begin{figure*}[t]
    \centering
    \includegraphics[scale=1]{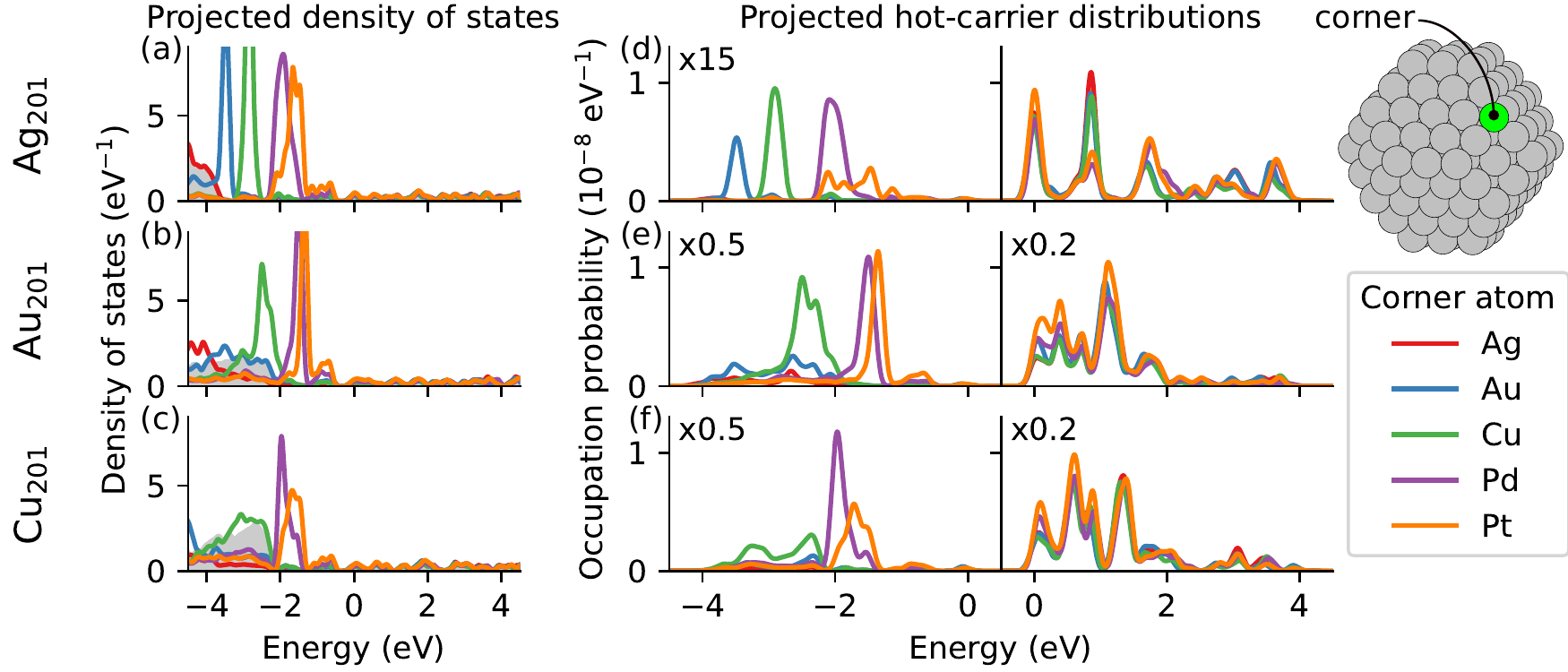}
    \caption{%
        (a-c) Density of states and (d-f) hot-carrier distributions projected to the corner atom in \ce{Ag201}, \ce{Au201}, and \ce{Cu201} NPs with the same corner atom being replaced with Ag, Au, Cu, Pd, or Pt.
        Note that the y-axis limits have a multiplier indicated in each panel.
        The gray shaded DOS corresponds to the total DOS in the plain NP divided by the number of atoms.
    }
    \label{fig:hcdist_corner}
\end{figure*}
%%%% FIGURE %%%%%%%%%%%%%%%%%%%%%%%%%%%%%%%%%%%%%%%%%%%%%%%%%%%%%%%%%%%%%%%%%%

%%%% FIGURE %%%%%%%%%%%%%%%%%%%%%%%%%%%%%%%%%%%%%%%%%%%%%%%%%%%%%%%%%%%%%%%%%%
\begin{figure}[h!]
    \centering
    \includegraphics[scale=1]{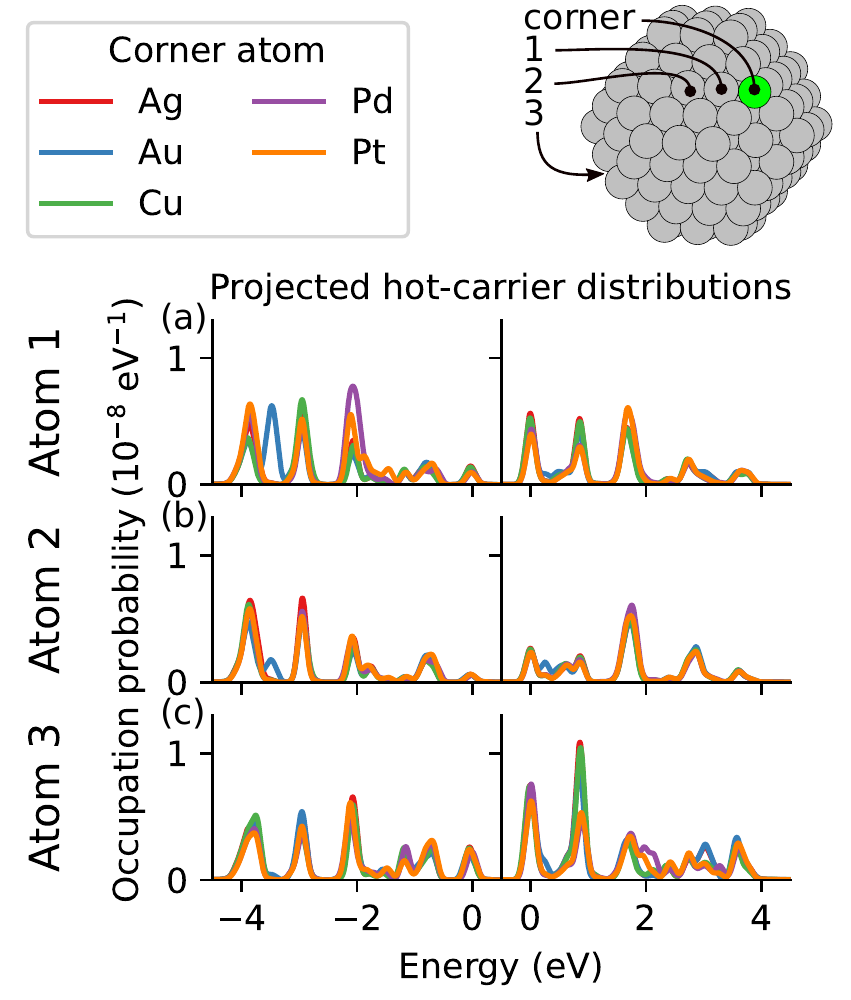}
    \caption{%
        (a-c) Hot-carrier distributions in \ce{Ag201} with the corner replaced with Ag, Au, Cu, Pd, or Pt projected to 1) adjacent atom, 2) nearby atom, and 3) opposite corner atom.
    }
    \label{fig:hcdist_corner_other_sites}
\end{figure}
%%%% FIGURE %%%%%%%%%%%%%%%%%%%%%%%%%%%%%%%%%%%%%%%%%%%%%%%%%%%%%%%%%%%%%%%%%%

The LSPR of the \ce{Ag201} particle is located at 3.77\,eV corresponding to a strong peak in the photoabsorption spectrum. In the larger \ce{Ag309} particle, the LSPR is shifted to slightly lower energies at 3.67\,eV (\autoref{fig:system}b).
\ce{Au201} and \ce{Cu201} NPs do not exhibit similarly strong LSPR peaks due to their earlier d-band and screening onset. \cite{CazDolRub00}

The plasmon formation and decay to hot-carriers in plain Ag NPs has already been discussed in detail in Ref.~\onlinecite{RosErhKui20}.
Introducing a single Ag, Au, Cu, Pd, or Pt atom as a dopant in the structure does not significantly change the photoabsorption strength (Supplementary Fig.~S2), the total density of states (DOS; Supplementary Fig.~S3), or the total hot-carrier distributions (Supplementary Fig.~S4) as can be expected from the low concentration of the dopant atoms.
However, the local electronic structure and plasmonic-catalytic properties are significantly affected by the single dopant atoms as will be discussed in the following.

The electronic structure of the metals Ag, Au, Cu, Pd, and Pt differ especially in terms of their d-electron bands. \cite{RahTibRos20}
The d-electron bands originate from the atomic d-states, and when a single dopant atom is embedded in the host NP, the local density of states projected onto the dopant atom exhibits strong localized states at different energies depending on the element (\autoref{fig:hcdist_corner}a-c).
The d-band onset of the plain \ce{Ag201} particle lies at $-3.8$\,eV, and Au, Cu, Pd, and Pt dopants in \ce{Ag201} exhibit localized states at $-3.5$\,eV, $-2.8$\,eV, $-1.9$\,eV, and $-1.6$\,eV, respectively (\autoref{fig:hcdist_corner}a).
These induced electronic states are very narrow, which indicates that the d-electron levels of the dopant atom do not interact strongly with the electronic states of the host NP.
This is especially clear for Au and Cu dopants in \ce{Ag201}, whereas Pd and Pt dopant show more hybridization with Ag NP states (\autoref{fig:hcdist_corner}a).

The changes in the local DOS have significant impact on the plasmonic-catalytic properties of the NPs;
the local DOS is strongly reflected in the hot-carrier distributions\cite{FojRosKui22} as seen in the dopant-atom-projected hot-hole distributions of \ce{Ag201} (\autoref{fig:hcdist_corner}d).
The hot-hole peaks occur at different energies depending on the dopant element, which suggests that the local catalytic properties could be tuned by choosing suitable dopant elements.
In contrast, the atom-projected hot-electron distributions (\autoref{fig:hcdist_corner}d) do not vary as strongly as those of plain \ce{Ag201} NP (Ag data in \autoref{fig:hcdist_corner}d) as they are dominated by delocalized sp-electrons with similar electronic structure.
This difference indicates that the electron states corresponding to the localized holes states are distributed over the NP.

%%%% FIGURE %%%%%%%%%%%%%%%%%%%%%%%%%%%%%%%%%%%%%%%%%%%%%%%%%%%%%%%%%%%%%%%%%%
\begin{figure*}[t]
    \centering
    \includegraphics[scale=1]{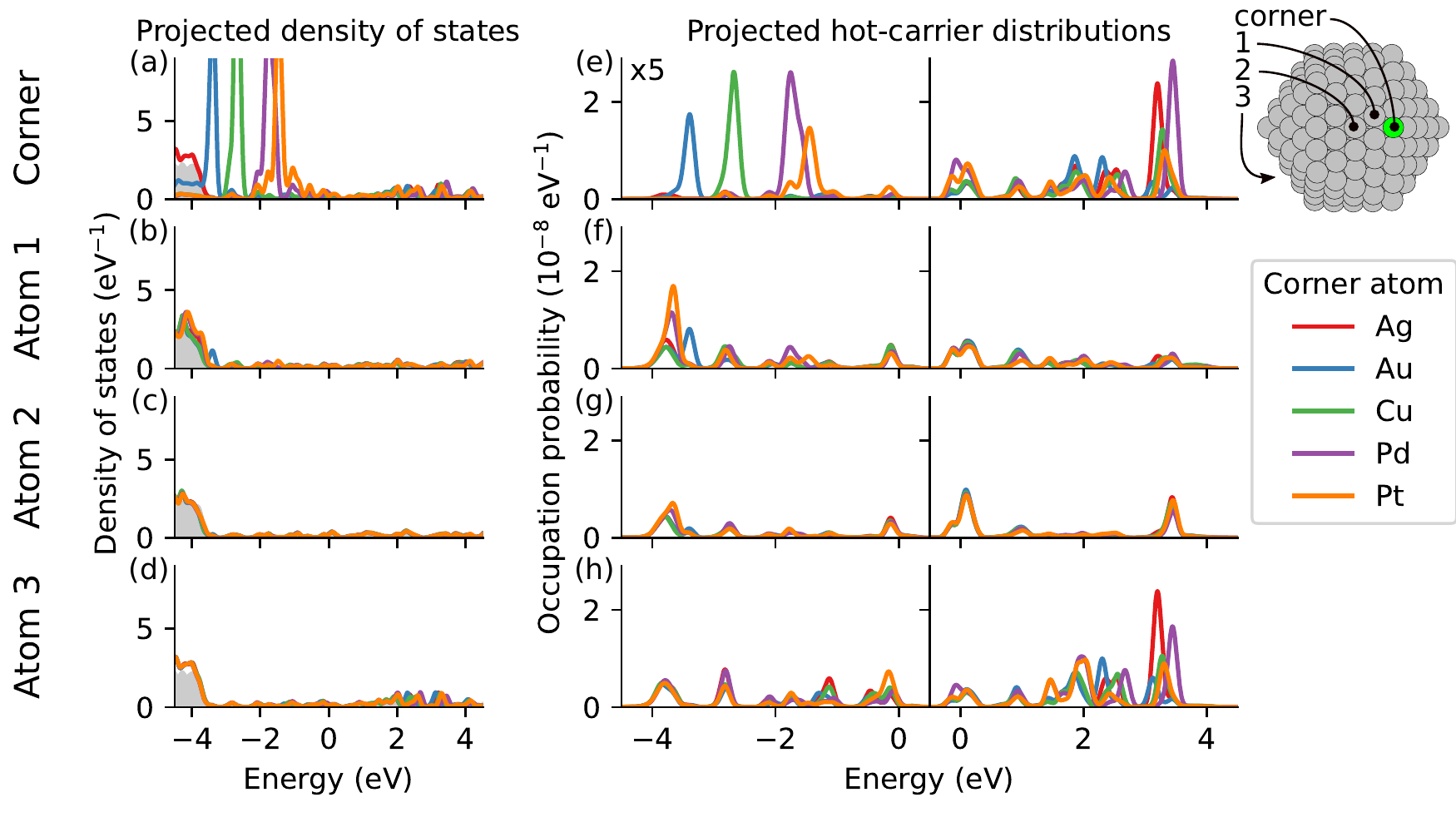}
    \caption{%
        (a-d) Density of states and (e-h) hot-carrier distributions projected to the corner and other atoms in \ce{Ag309} NP with the corner atom replaced with Ag, Au, Cu, Pd, or Pt.
        Note that the y-axis limits have a multiplier indicated in each panel.
        The gray shaded DOS corresponds to the total DOS in the plain NP divided by the number of atoms.
    }
    \label{fig:hcdist_Ag309}
\end{figure*}
%%%% FIGURE %%%%%%%%%%%%%%%%%%%%%%%%%%%%%%%%%%%%%%%%%%%%%%%%%%%%%%%%%%%%%%%%%%

It is instructive to inspect the ordering of the dopant d-electron levels in the \ce{Ag201} host NP: Ag $<$ Au $<$ Cu $<$ Pd $<$ Pt (\autoref{fig:hcdist_corner}a,d).
Following this ordering in the other host NPs, we note that for \ce{Au201}, the d states of the Ag dopant are pushed below the d-band onset of Au and the atom-projected hot-hole distribution is low(\autoref{fig:hcdist_corner}b,e).
Correspondingly, for the \ce{Cu201} host NP, the d states of both Ag and Au are pushed below the d-band onset of Cu and these dopants again do not contribute to the generation of hot holes (\autoref{fig:hcdist_corner}c,f).
This demonstrates the importance of understanding the electronic structure of the combined system for tuning its hot-carrier distributions.

Qualitatively similarly peaked DOS and HC distributions are also observed for dopant atoms on a \{111\} NP facet (Supplementary Fig.~S5).
However, the local environment of the dopant atom on the denser \{111\} facet differs from that of the corner site we have inspected so far.
This causes quantitative differences especially in the case of Au on \ce{Ag201} with a weaker HC distribution and in the case of Pt in \ce{Ag201} with a more discrete HC distribution.

The atom-local nature of the peaked hot-hole distributions is further confirmed by analyzing atom-projected quantities at other atomic sites in the system (shown for \ce{Ag201} host NP in \autoref{fig:hcdist_corner_other_sites}).
The hot-hole distributions projected on the atom adjacent to the dopant corner atom in \ce{Ag201} exhibit only minor changes in comparison to the plain \ce{Ag201} HC distributions (\autoref{fig:hcdist_corner_other_sites}a).
The differences become negligible at the next-nearest atom on a \{111\} facet (\autoref{fig:hcdist_corner_other_sites}b).
Only the discrete state of the Au dopant in \ce{Ag201} appears more visible as its energy is different from the other states in the distributions.
Interestingly, while the hot-hole distributions at the corner atom on the opposite site of the NP (\autoref{fig:hcdist_corner_other_sites}c) are unaffected by the dopant, the hot-electron distributions exhibit changes although the distance to the dopant atom is maximal.
These changes are likely due to the disruptions in the delocalized electronic states and super-atomic shell structure \cite{Hak16} caused by the dopant atom in the otherwise highly symmetric NP.
Furthermore, the hot-electron distribution at the opposite corner atom resembles closely the distribution at the dopant corner atom (compare \autoref{fig:hcdist_corner_other_sites}c and \autoref{fig:hcdist_corner}d), which indicates that the changes in the hot-electron distributions are of this same origin and present throughout the particle.

\begin{table}[t]
\caption{Energies of the NP structures with a single dopant atom placed in different subsurface layers. Layer 0 corresponds to the atom on the surface at a \{111\} facet site and the subsequent layers are deeper in the particle, layer 3 being at the center of the NP. Energies are in meV relative to the minimum energy structure (0 meV).}
\label{tab:energies}
\centering
\begin{tabular}{c|rrrr|rrrr|rrrr}
\hline\hline
\multirow{2}{*}{Layer} & \multicolumn{4}{c|}{\ce{Ag201}} & \multicolumn{4}{c|}{\ce{Au201}} & \multicolumn{4}{c}{\ce{Cu201}} \\
%        & Ag       & Ag       & Ag       & Ag       & Au       & Au       & Au       & Au       & Cu       & Cu       & Cu       & Cu       \\
         & Au       & Cu       & Pd       & Pt       & Ag       & Cu       & Pd       & Pt       & Ag       & Au       & Pd       & Pt       \\
\hline
       0 &        0 &      135 &      381 &      371 &      184 &      283 &      493 &      389 &        0 &        0 &        0 &        0 \\
       1 &       97 &        0 &       79 &      121 &       42 &       18 &       67 &       90 &      493 &      610 &       98 &      201 \\
       2 &       69 &       13 &        0 &        0 &       63 &       26 &        7 &        0 &      524 &      571 &      127 &      150 \\
       3 &      112 &       34 &       39 &      111 &        0 &        0 &        0 &       17 &      581 &      721 &      465 &      674 \\
\hline\hline
\end{tabular}
\end{table}

Icosahedral particles exhibit pronounced hot-electron generation at their corner sites. \cite{RosErhKui20}
Replacing a corner atom with a dopant in an icosahedral \ce{Ag309} particle results in a qualitatively similar effects as in the octahedral \ce{Ag201} particle considered earlier.
The dopant corner site exhibits distinct localized electronic states above the d-band onset of the Ag host (\autoref{fig:hcdist_Ag309}a), which results in peaked hot-hole distribution at the dopant site (\autoref{fig:hcdist_Ag309}e).
The details of the hot-hole distributions are, however, different between \ce{Ag309} and \ce{Ag201}, especially for the Pt dopant that shows a spread-out hot-hole distribution in \ce{Ag201}, but a more discrete distribution in \ce{Ag309}.
The qualitative trends and the ordering of the Au, Cu, Pd, and Pt dopant atom states remain similar between \ce{Ag201} and \ce{Ag309}, indicating their likely generality.
However, more data would be needed to draw robust conclusions.
Analyzing the hot-carrier distributions at other atomic sites in \ce{Ag309} (\autoref{fig:hcdist_Ag309}f-h) reveals the atomic locality of the dopant-induced changes in the hot-carrier distributions in analogy to \ce{Ag201}.
Furthermore, the \ce{Ag309} NP confirms the previously noted sensitivity of the super-atomic delocalized electronic states to the single dopant atom: The changes in the hot-electron distributions at the corner dopant atom (\autoref{fig:hcdist_Ag309}e) resembles the changes at the opposite corner atom (\autoref{fig:hcdist_Ag309}h).

Finally, to estimate thermodynamic stability of the modeled structures, we evaluated the total energy of the NP with a dopant atom at different subsurface layers (\autoref{tab:energies}).
The considered dopant atoms are energetically most stable on the surface of \ce{Cu201} NP.
Conversely, for \ce{Ag201} and \ce{Au201}, the dopant atom is most stable below the surface, except for Au in \ce{Ag201}.
While most of the structures considered are not global minima, they are still in local minimum energy configurations and might be stable enough kinetically.

\section{Conclusions and Outlook}
\label{sec:conclusions}

In this study, we investigated plasmonic hot-carrier generation in Ag, Au, and Cu nanoparticles with single Ag, Au, Cu, Pd, or Pt dopant atoms.
The performed DFT and TDDFT calculations indicate that these dopant atoms exhibit discrete occupied atom-localized states in the host nanoparticle, which result in significantly peaked local hot-hole distributions while the distribution of hot-electrons is less affected, and the photoabsorption spectrum of the nanoparticle remain mostly unchanged.
The energy of the generated hot holes depends on the dopant element across different host nanoparticles (Ag, Au, Cu), different nanoparticle sizes (octahedral 201-atom and icosahedral 309-atom Ag nanoparticle), and different dopant atom locations (corner and facet sites).
This suggests that the local hot-hole generation could be optimized by choosing suitable dopant elements.
In the context of plasmonic catalysis, this tunability could be further utilized for designing selective catalyst by matching the hot-carrier distribution with desired energy levels.
It would be interesting to extend the presented analysis by including target molecules explicitly in the calculations to take into account orbital hybridization effects and level alignment \cite{FojRosKui22}.

\section*{Software used}

The GPAW package \cite{MorHanJac05, EnkRosMor10} with localized basis sets \cite{LarVanMor09} and the LCAO-RT-TDDFT implementation \cite{KuiSakRos15} was used for the TDDFT calculations.
The Libxc library \cite{LehSteOli18} was used exchange-correlation functionals.
The ASE library \cite{LarMorBlo17} was used for constructing, manipulating, and relaxing atomic structures.
The NumPy \cite{HarMilvan20}, SciPy \cite{VirGomOli20} and Matplotlib \cite{Hun07} Python packages were used for processing and plotting data.
The Snakemake \cite{MolJabLet21} package was used for managing the calculation workflow.

\begin{acknowledgments}
T.P.R. acknowledges funding from the Academy of Finland under grant No~332429.
The calculations were enabled by the generous computational resources provided by CSC -- IT Center for Science, Finland and by the Aalto Science-IT project, Aalto University School of Science.
\end{acknowledgments}

\bibliography{lit.bib}

\end{document}